\documentclass[print,12pt]{revtex4}
\usepackage{graphicx,epsfig}

\begin{document}

\title{Spherical Collapse with Dark Energy}

\author{Irit Maor}
 \affiliation{CERCA, Department of Physics,
Case Western Reserve University, 10900 Euclid Ave, Cleveland, OH
44106-7079 USA}
 \email{irit.maor@case.edu}

\date{\today}

\begin{abstract}
I discuss the work of Maor and Lahav \cite{ml}, in which the
inclusion of dark energy into the spherical collapse formalism is
reviewed. Adopting a phenomenological approach, I consider the
consequences of - a) allowing the dark energy to cluster, and, b)
including the dark energy in the virialization process. Both of
these issues affect the final state of the system in a fundamental
way. The results suggest a potentially differentiating signature
between a true cosmological constant and a dynamic form of dark
energy. This signature is unique in the sense that it does not
depend on a measurement of the value of the equation of state of
dark energy.
\end{abstract}

\maketitle
\section{Introduction}

One of the outstanding issues of cosmology is dark energy. The
primary question is whether dark energy is a cosmological
constant, or is it dynamical. In order to use inhomogeneity
studies to probe dark energy, it is essential that we understand
how the presence of dark energy affects the evolution of
overdensities. Adopting a phenomenological approach, the aim of
this work is to consider what are the effects on the evolution of
inhomogeneities if the dark energy clusters, or, alternatively, if
it participates in the virialization process. It is based on the
work of Maor and Lahav \cite{ml}.

A fundamental tool in the analysis of inhomogeneities is the
spherical collapse formalism, which dates back to Gunn and Gott
\cite{gg}. It describes how a small spherical patch of homogeneous
over-density decouples from the expansion of the universe, slows
down, and eventually turns around and collapses. It is assumed
that the collapse is not complete, thus it does not lead into a
singularity. Instead, the system eventually virializes and
stabilizes, having a finite size. The definition of the moment of
virialization depends on energy considerations. The top hat
spherical collapse is incorporated, for example, in the
Press-Schecter \cite{ps} formalism. It is, therefore, widely used
in present day interpretation of data sets.

The generalization of the spherical collapse formalism to include
additional forms of energy has been subject to numerous studies
\cite{llpr,ws,is,wk,bw,zg,mb,hb,wang,ml}. Lahav {\it et al}
\cite{llpr} generalized the formalism to a universe composed of
ordinary matter and a cosmological constant. Wang and Steinhardt
\cite{ws} included Quintessence with a constant or a slowly
varying equation of state. Battye and Weller \cite{bw} included
Quintessence in a different manner than Wang and Steinhardt,
taking into account its pressure. Mota and Van de Bruck \cite{mb}
considered spherical collapse for different potentials of the
Quintessence field, and checked what happens when one relaxes the
common assumption that the Quintessence field does not cluster on
the relevant scales. Maor and Lahav \cite{ml} considered both
clustering and homogeneous dark energy, and examined what are the
effects if dark energy participates in the virialization process.
They pointed out a source of energy non conservation in the case
where dark energy is kept homogeneous, and suggested how to
incorporate this energy non conservation. Wang \cite{wang}
considered another source of energy non-conservation, due to the
fact that a homogeneous dark energy acts as a time-dependent, and
hence nonconservative force.

The structure of this paper is as follows. Section \ref{sc}
reviews the basics of the spherical collapse formalism. The
procedure by which we define virialization of an overdensity is
reviewed in section \ref{virialization}. For non clustering dark
energy which is not a cosmological constant there are some
problems regarding energy conservation, which are discussed in
section \ref{nec}. Section \ref{results} presents some results,
and section \ref{conclusions} is dedicated to concluding remarks.

\section{Spherical collapse}
\label{sc}

We take the background cosmology to be a flat FRW universe with
two energy components. One is non relativistic dust $\rho_m$ with
pressure $p_m=0$ (for the sake of this discussion it is
unimportant if this component is luminous or not). The second
component is the dark energy, modelled as a perfect fluid with
pressure $p_Q=w\rho_Q$, $w$ being the (constant) equation of
state. The equations governing the background evolution are then
\begin{eqnarray}
 & \left(\frac{\dot a}{a}\right)^2 = \frac{8\pi G}{3}\left(
      \frac{}{}\rho_m+\rho_Q \right) & \\
 & \frac{\ddot a}{a} = -\frac{4\pi G}{3}\left(\frac{}{}\rho_{m}+
      \left(1+3w \right)\rho_{Q}\right) & \\
 & \dot\rho_m+3\left(\frac{\dot a}{a}\right)\rho_m = 0 & \\
 & \dot\rho_Q+3(1+w)\left(\frac{\dot a}{a}\right)\rho_Q = 0 ~, &
\end{eqnarray}
where $a$ is the global scale factor.

Within such a universe, we assume that there is a spherical
perturbation in the matter density, with a flat (top hat) profile.
$\rho_{mc}$ denotes the matter density within the perturbation. We
assume that the initial perturbation is in the matter field only,
though we will allow non-homogeneity to develop for the additional
fluid. Following the spherical collapse formalism, the equations
governing the evolution of the overdensity are similar to those of
the background, with the global scale factor $a$ replaced with the
local scale factor $R$. The flatness condition is not held,
because of the perturbation in the matter,
\begin{eqnarray}
 & \frac{\ddot R}{R} = -\frac{4\pi G}{3}\left(\frac{}{}\rho_{mc}+
      \left(1+3w \right)\rho_{Qc}\right) & \\
 & \dot\rho_{mc}+3\left(\frac{\dot R}{R}\right)\rho_{mc} = 0 & \\
 & \dot\rho_{Qc}+3(1+w)\left(\frac{\dot R}{R}\right)\rho_{Qc} =
      \gamma\Gamma \label{qc} ~, &
\end{eqnarray}
with
\begin{eqnarray}
 & \Gamma  =  3(1+w)\left(\frac{\dot R}{R}-
      \frac{\dot a}{a}\right)\rho_{Qc} & \\
 & 0\leq \gamma \leq  1 ~. &
\end{eqnarray}
The form of equation (\ref{qc}) allows us to move, in a continuous
way, between two cases of interest. The first case is where the
dark energy is kept homogeneous, by choosing $\gamma=1$. Equation
(\ref{qc}) then reads $\dot\rho_{Qc}+3(1+w)\left(\frac{\dot
a}{a}\right)\rho_{Qc} =0$, and, therefore, the evolution of the
dark energy within the perturbation is similar to that of the dark
energy in the background, $\rho_{Qc}=\rho_Q$. The second case of
interest is where the dark energy is allowed to follow the local
scale factor and fully cluster. This is done by taking $\gamma=0$,
in which case equation (\ref{qc}) reads
$\dot\rho_{Qc}+3(1+w)\left(\frac{\dot R}{R}\right)\rho_{Qc} =0$.
Thus one can think of $\gamma$ as a clustering parameter. This new
parameter introduces a new scale to the problem, which defines the
clustering rate of the dark energy. $\gamma\Gamma$ is the rate in
which the perturbation loses energy to the background, due to the
$Q$ field. The system is conservative when $\gamma=0$ and the $Q$
field is allowed to fully cluster, or when $\Gamma=0$. The latter
case corresponds to $w=-1$. In all other cases, the system loses
energy to the background, an issue which  will be addressed later
on.\\

The clustering properties of dark energy are the subject of recent
debate. Even though Caldwell {\it{et al}} \cite{c} have shown that
Quintessence cannot be perfectly smooth, it is assumed that the
clustering is negligible on scales less than $100~Mpc$. It is,
therefore, a common practice to keep the Quintessence homogeneous
during the evolution of the system. On the other hand, one should
bear in mind that every positive energy component other than the
cosmological constant is capable of clustering, the question is at
what rate. Additionally, we are looking at the evolution of the
perturbation well beyond the linear regime, where the reaction of
the dark energy to the local metric is unclear. Clustering of dark
energy is particularly well motivated for models in which dark
energy is coupled, in some form, to the matter \cite{mb,qc}. The
resolution of this debate cannot be found in the prescription of
the spherical collapse, Moreover, a top hat profile for the dark
energy is not a stable configuration. This work does not attempt
to provide an answer to whether and how the dark energy clusters,
but rather explores the consequences of such a scenario. A more
fundamental treatment of the clustering properties of dark energy
is needed, and is a subject of an ongoing investigation
\cite{wip}. ''

The above equations, supplied with the appropriate initial
conditions, can now be solved. Following the mathematical solution
of the perturbations leads to a decoupling of the perturbation
from the background. The local scale factor $R$ evolves in a
slower fashion than the global scale factor $a$, reaches its
maximal size $R_{ta}$ at turnaround, and then the system begins to
collapse. Following the mathematical solution all the way through
leads to a singularity.

\section{Virialization}
\label{virialization}

Even though the mathematical solution of the spherical collapse
equations gives a point singularity as the final state of the
system, we know that, physically, objects go through a
virialization process, and stabilize towards a finite size.
Virialization is not `built in' into the spherical collapse model
(see though \cite{pad}), and the common practice is to
{\it{define}} the virialization radius as the radius at which the
virial theorem holds, and the kinetic energy $T$ is related to the
potential energy $U$ by
\begin{eqnarray}
  & T_{vir} = \left(\frac{R}{2}\frac{\partial U}{\partial
      R}\right)_{vir} ~. &
\end{eqnarray}
Using energy conservation between virialization and turnaround
(where $T_{ta}=0$) gives
\begin{eqnarray}
 & \left(U+\frac{R}{2}\frac{\partial U}
      {\partial R}\right)_{vir} = U_{ta} ~. & \label{vir}
\end{eqnarray}
Equation (\ref{vir}) defines $R_{vir}$.

The fact that the details of the virialization process are
bypassed by the above procedure is useful because these details
are complex and not fully understood. On the other hand, this
means that this procedure cannot provide us with information about
how the dark energy behaves during the virialization process. The
above energy budget should be applied to {\it{those components
that virialize}}. If only the matter virializes, equation
(\ref{vir}) becomes
\begin{eqnarray}
 & \left(U(\rho_{mc})+\frac{R}{2}\frac{\partial U(\rho_{mc})}
      {\partial R}\right)_{vir} = U(\rho_{mc})_{ta} ~, &
\end{eqnarray}
and if the whole system virializes as a whole - both the matter
and the $Q$ field, equation (\ref{vir}) should then read
\begin{eqnarray}
 & \left(U(\rho_{tot})+\frac{R}{2}\frac{\partial U(\rho_{tot})}
      {\partial R}\right)_{vir} = U(\rho_{tot})_{ta} ~, &
      \label{tot}
\end{eqnarray}
with $\rho_{tot}=\rho_{mc}+\rho_{Qc}$. As will be shown in the
next section, the choice which of the components actually
virializes make a significant difference. \\

Whether the dark energy participates in the virialization is an
open question. In principle, any energy component with non
vanishing kinetic energy is capable of virializing, given enough
time. Once again, our aim is not to settle the issue of the
virialization properties of the dark energy, but rather consider
and compare the consequences of both options. While it seems
reasonable to connect between the clustering and the virializing
properties of dark energy, this may lead to some problems, because
clustering is a continuous property, while the virialization is
not. We will, therefore, consider various degrees of clustering,
regardless of whether the dark energy virializes or not.

\section{Energy non conservation}
\label{nec}

Equation (\ref{tot}) assumes energy conservation for the
virializing component(s) between turnaround and the time of
virialization. As was mentioned earlier, the $Q$ field loses
energy to the background if $\gamma\Gamma\neq 0$, that is if the
dark energy is {\it{not}} a cosmological constant, but is
nonetheless not allowed to follow the local metric and cluster. If
such a $Q$ field participates in the virialization, then equation
(\ref{tot}) needs to be corrected, in order to account for the
energy lost. \\

The continuity equation for the energy losing field is
\begin{eqnarray}
 & \dot\rho_{Qc}+3(1+w)\left(\frac{\dot R}{R}\right)\rho_{Qc} =
      \gamma\Gamma ~. &
\end{eqnarray}
We can imagine a field with the same equation of state that
{\it{does}} conserve energy, $\tilde{\rho}_{Qc}$. The continuity
equation for this energy-conserving field is then
\begin{eqnarray}
 & \dot{\tilde\rho}_{Qc}+3(1+w)\left(\frac{\dot R}{R}\right)\tilde{\rho}_{Qc} =
      0 ~. &
\end{eqnarray}
If we chose the same initial conditions for the real field and for
the energy-conserving field at turnaround,
$\tilde{\rho}_{Qc}(t_{ta})=\rho_{Qc}(t_{ta})$, then by
construction, the amount of the energy which was lost at a later
time is
\begin{eqnarray}
 & \Delta U \equiv \tilde U-U ~, &
\end{eqnarray}
where $\tilde{U}(\rho_{Qc})\equiv U(\tilde{\rho}_{Qc})$ is the
potential energy function for the energy-conserving field. The
correction to equation (\ref{tot}) is then
\begin{eqnarray}
 & \left( U+\Delta U+\frac{R}{2}\frac{\partial U}{\partial R}
    \right)_{vir} =
    \left(\tilde U+\frac{R}{2}\frac{\partial U}{\partial
    R}\right)_{vir}
    = U_{ta} ~. & \label{new}
\end{eqnarray}
As will be shown in the next section, this introduces a small
quantitative correction to the final state of the system. \\

A different source of energy non conservation (ENC) is when the
$Q$ field is kept homogeneous (or does not fully cluster,
$\gamma\neq 0$), and does not participate in the virialization.
The virializing component then feels a time dependent, non
conservative force. This energy non conservation was addressed in
\cite{wang}.

\section{Results}
\label{results}

\begin{figure}
  \begin{center}
    \epsfig{file=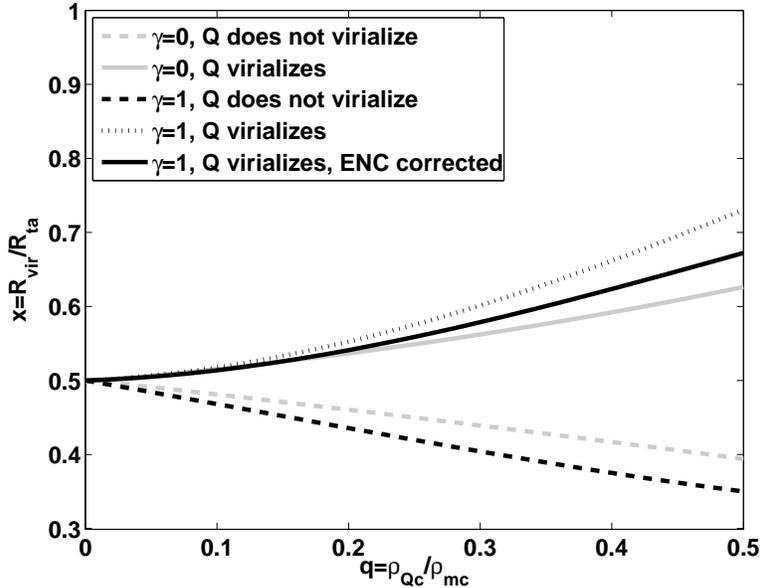,height=80mm}
  \end{center}
  \caption{The ratio of final to turnaround radii, $x=R_{vir}/R_{ta}$,
  as a function of $q=\rho_{Qc}/\rho_{mc}$ at turnaround, for
  Quintessence with a constant equation of state $w=-0.8$.
  \label{w8}}
\end{figure}

Figure \ref{w8} gives a summary of the different choices one can
make when applying the spherical collapse formalism to a cosmology
with dark energy. It shows the dependence of the final size of an
overdense system relative to its maximal size, as a function of
the strength of the dark energy at the time of turnaround,
$q=(\rho_{Qc}/\rho_{mc})_{ta}$. For all curves the equation of
state of the dark energy is $w=-0.8$. The grey lines are for fully
clustering dark energy, while the black lines are for the
homogeneous case. Dashed lines follow the behavior of the system
when only the matter component virializes, and the solid lines are
for the case when the complete system virializes as a whole,
including the dark energy component. Comparison of the dotted line
with the solid black line shows the effect of the ENC correction.
As can be seen, all cases converge to $x=1/2$ as $q\rightarrow 0$,
which is the analytical result for a universe made strictly of
matter. In general, one can say that if only the matter
virializes, the presence of dark energy creates bound objects
which are smaller and denser. Allowing the dark energy to
participate in the virialization results in larger, less dense
objects. This effect is enhanced if the dark energy is kept
homogeneous. Correcting for the energy loss of a homogeneous dark
energy that virializes slightly weakens the effect, but does not
introduce any qualitatively new behavior. \\

\begin{figure}
  \begin{center}
    \epsfig{file=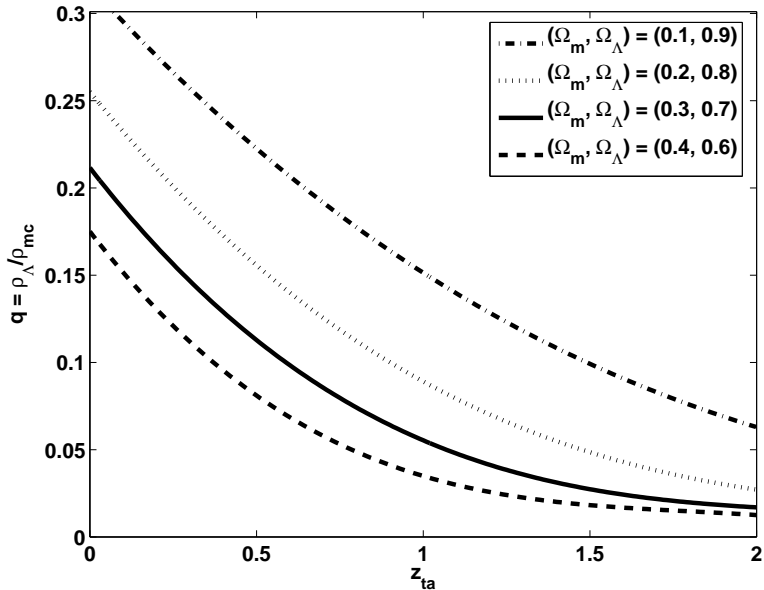,height=80mm}
  \end{center}
  \caption{$q=\rho_{\Lambda}/\rho_{mc}$ as a function of the
  turnaround redshift $z_{ta}$, for various values of $\Omega_m$
  and $\Omega_{\Lambda}$.
  \label{qz}}
\end{figure}

The figure shows how the various cases behave as a function of
$q$, which is, in turn, dependent on the cosmological model, as
well as on the time in which the turnaround happens. Figure
\ref{qz} shows the dependence of $q$ as a function of the
turnaround redshift, for various $\Lambda CDM$ cosmologies. As the
figure shows, typical $q$ values of cosmological interest are less
than $0.3$.

\begin{figure}
  \begin{center}
    \epsfig{file=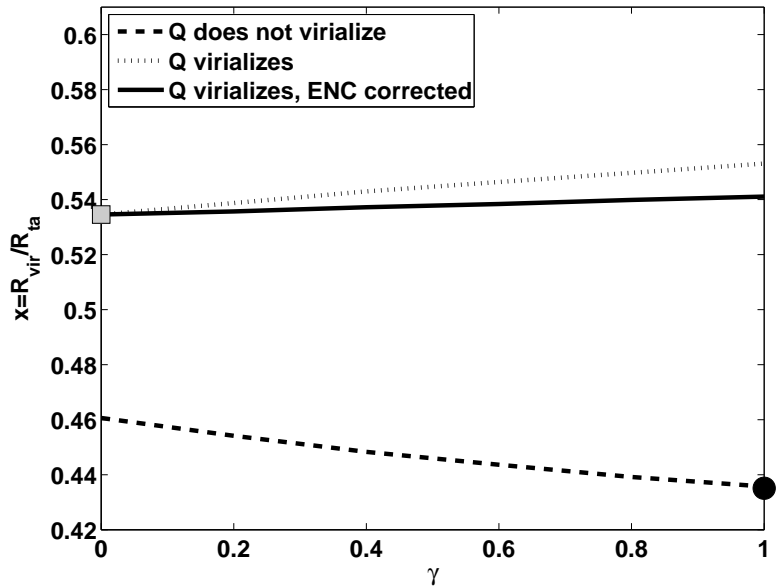,height=80mm}
  \end{center}
  \caption{$x=R_{vir}/R_{ta}$ as a function of the clustering
  parameter $\gamma$, for $w=-0.8$ and $q=0.2$. $\gamma=0$ describes
  the case of a fully clustering $Q$ field, and $\gamma=1$ is the
  case of a homogeneous $Q$, allowing only the matter component
  to cluster. For $\gamma=0$, taking the dark energy into the
  virialization is highly plausible, (see square on left).
  If one assumes that only the matter component virializes
  for $\gamma=1$ (see circle on right), it is unclear how to
  extrapolate in a smooth way between the two cases.
  This will produce a discontinuity in the transition
  from the `clustering' to the `non-clustering' behavior.
  \label{gamma}}
\end{figure}

A better understanding of what the virialization process consists
of is needed in order to decide whether dark energy participates
in it, and here one can only speculate. The most motivated case
for allowing the dark energy to virialize is when it clusters, for
the clustering can be interpreted as a sign that the dark energy
feels and responds to the local interactions, and possibly also to
those that lead to virialization. A natural approach, therefore,
would be to associate between the clustering and the virialization
processes. This approach poses a problem though, illustrated in
figure \ref{gamma}. The figure shows the solutions of $x$ as a
function of $\gamma$, with fixed $w=-0.8$ and $q=0.2$. The circle
on the right is the Wang and Steinhardt's result when the
quintessence is kept completely homogeneous, and only the matter
component virializes. The square on the left is the result when
the dark energy fully clusters, and both the matter and the dark
energy virialize. One would expect the transition in the behavior
of the system along $\gamma$ to be smooth. Allowing the dark
energy to virialize for the clustering case, $\gamma=0$, and
keeping it out of the virialization process when $\gamma=1$,
raises the question of how one should extrapolate smoothly between
the two cases. As figure \ref{gamma} suggests, there will be a
discontinuity. \\

\begin{figure}
  \begin{center}
    \epsfig{file=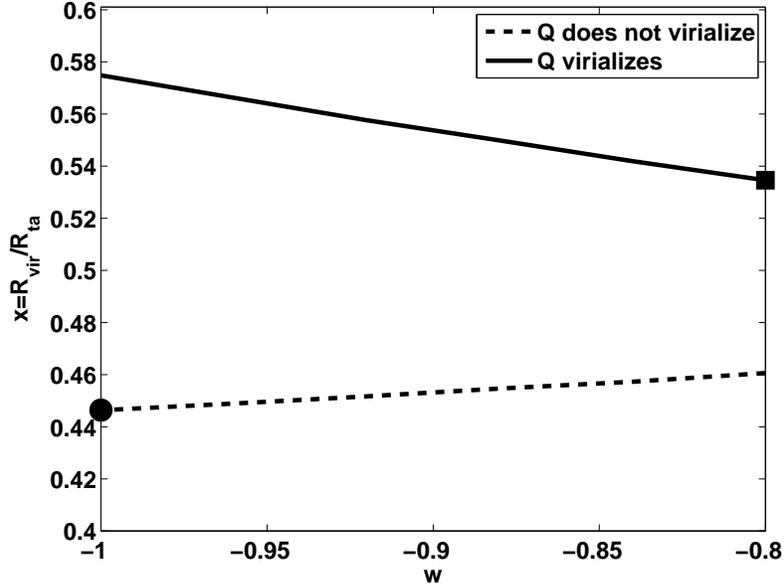,height=80mm}
  \end{center}
  \caption{$x=R_{vir}/R_{ta}$ as a function of the equation of state
  $w$, for $q=0.2$ and $\gamma=0$. The dashed line is the ratio when
  the matter alone virializes, and the solid is for the case where
  the whole system virializes. The circle on the left is Lahav {\it{et
  al}}'s solution for the cosmological constant \cite{llpr}. The square
  on the right is an example of a clustered quintessence, where we
  expect to take into account the whole system in the virialization.
  The figure suggests we should expect a discontinuity in the
  behavior of quintessence fields and a true cosmological
  constant.
  \label{w0}}
\end{figure}

Another issue which deserves special attention is the limit of the
cosmological constant, $w\rightarrow -1$. For $w=-1$ the
clustering parameter $\gamma$ plays no role, because the question
whether such a fluid is allowed to cluster ($\gamma=0$) or not
($\gamma=1$) is rather abstract. It stays homogeneous in any case,
because of its equation of state, $w_{\Lambda}=-1$ (which leads to
$\Gamma=0$). Accordingly, energy is automatically conserved. An
equivalent of figure \ref{w8} drawn for $w=-1$ will show that the
grey and the black dashed lines coincide, as well as the grey and
the black solid lines. The dotted line will coincide with the
solid ones as well, because $\Delta U=0$. The difference between
dark energy which does or doesn't virialize is still evident
though. Again, one should consider the plausibility of the two
solutions. If one considers the cosmological constant as a true
constant of Nature, $\rho_{\Lambda}=\Lambda/(8 \pi G)$, it is hard
to imagine it participating in the dynamics that lead to
virialization, as it is a true constant. In this case, one could
categorically say that the right procedure is to look at the
virialization of the matter fluid only, following the work of
Lahav {\it{et al}} \cite{llpr} (see circle on left in figure
\ref{w0}). The sole effect of the cosmological constant, then, is
to modify the potential that the matter feels, through the
background expansion.

If, on the other hand, one considers the origin of a perfect fluid
with $w\approx -1$ as a special case of quintessence, which is
indistinguishable from a cosmological constant, it is reasonable
to expect continuity in the behavior of the system as one slowly
changes the value of $w$ toward $-1$. In other words, if the
physical interpretation of the fluid with $w\approx -1$ is of a
dynamical field that {\it{mimics}} a constant, the idea of
including it in the dynamics of the system has a physical meaning.

The result, then, is that we possibly have a signature
differentiating between a cosmological constant which is a true
constant, and a different entity which {\it{mimics}} a constant.
This point is shown in figure \ref{w0}. The figure shows $x$ as a
function of $w$, with fixed $q=0.2$ and $\gamma=0$. The dashed
line shows how $x$ depends on $w$ when only the matter virializes.
The circle on the left is Lahav {\it{et al}}'s solution for the
cosmological constant. The solid line shows how the system behaves
when both the matter and the dark energy virialize. The square on
the right is an example of a clustered dark energy, where we
expect to take into account the whole system in the virialization.
As with figure \ref{gamma}, there is a suggested discontinuity,
but here one can associate the discontinuity with a clear physical
meaning: a true cosmological constant is not on the continuum of
perfect fluids with general $w$, as its physical behavior is
different. An observational detection of virialized objects with
$R_{vir}>R_{ta}/2$ would be a strong evidence against a
cosmological constant which is a true constant, regardless of the
measured value of the equation of state.

\section{Conclusions}
\label{conclusions}

The inclusion of dark energy into the spherical collapse formalism
was reviewed. As the spherical collapse formalism is not a
calculation from first principles, it does not provide information
about the behavior of the dark energy during the evolution of an
overdensity. Specifically, one needs to decide whether to allow
the dark energy to cluster, and whether one should include it in
the virialization process. The consequences of these choices were
examined and are summarized in figure \ref{w8}. An additional
issue which was addressed here is the energy non-conservation,
which arises when the dark energy is kept homogeneous and is part
of the virialization.

The primary result suggests the possibility of an observational
signature differentiating between a true cosmological constant,
and a dynamical form of dark energy that mimics a constant. An
observational detection of virialized objects with
$R_{vir}>R_{ta}/2$ would be a strong evidence against a
cosmological constant which is a true constant. This signature is
unique in the sense that it does not depend on measuring the value
of the equation of state of dark energy, contrary to most existing
probes of dark energy. In order to discuss this signature in more
definite and quantitative terms, a better understanding of the
dark energy behavior during the evolution of overdensities is
required. The clustering properties of dark energy \cite{wip}, as
well as a more detailed understanding of how systems virialize,
need to be further explored. This information cannot be found in
numerical simulations. The incorporation of dark energy into
simulations is done by modifying the background expansion. It
means that by construction we will not be able to see clustering
or virialization of dark energy. Thus another direction to pursue
is how to improve the incorporation of dark energy into
simulations.

These directions of future research are particularly important,
because observational evidence seems to point that even if the
dark energy is in essence dynamical, it is doing a very good
imitation of a cosmological constant. As the implications of the
two options are so dramatically different, it is worth exhausting
all possibilities we can think of to distinguish between a
cosmological constant and dynamic dark energy. This is an exiting
challenge.

\begin{acknowledgements}

It is a pleasure to thank Prof.~Ofer Lahav for an enjoyable and
stimulating collaboration which is at the base of this review.
``Peyresq Physics 10" workshop provided an enjoyable and intimate
set-up for stimulating discussions about gravitation and
cosmology. I would like to thank the organizers, and in particular
Prof.~Edgard Gunzig, for the invitation to participate, and for
the efforts that went into the superb organization of the
workshop. I acknowledge the support of DOE and of NASA.

\end{acknowledgements}


\end{document}